\documentclass[prl,citeautoscript, reprint, amsmath, amssymb]{revtex4-2}

\usepackage{bm}
\usepackage{amsmath}
\usepackage{mathrsfs}
\usepackage{graphicx}
\usepackage{dcolumn}
\usepackage{bm}
\usepackage{lipsum}
\usepackage{gensymb}
\usepackage{esvect} 
\usepackage{balance}
\usepackage{float}
\usepackage{makecell}
\usepackage{booktabs}
\usepackage{natbib}

\usepackage[utf8]{inputenc}
\usepackage[english]{babel}

\usepackage{hyperref}
\hypersetup{ 
colorlinks=true,
linkcolor=blue,
filecolor=magenta,      
urlcolor=cyan,
pdftitle={Overleaf Example},
pdfpagemode=FullScreen,
citecolor=,
}

\newcommand{\rpm}{\sbox0{$1$}\sbox2{$\scriptstyle\pm$}\raise\dimexpr(\ht0-\ht2)/2\relax\box2 }

\begin{document}

\preprint{JRN/123-ABC}

\title{A versatile generalized digital twin for Electron Microscopy}

\author{Thomas W. Pfeifer}
\email{pfeifertw@ornl.gov}
\affiliation{Center for Nanophase Materials Sciences, Oak Ridge National Laboratory, Oak Ridge,  Tennessee 37830, USA}

\author{Andrew R. Lupini}
\email{arl1000@ornl.gov}
\affiliation{Center for Nanophase Materials Sciences, Oak Ridge National Laboratory, Oak Ridge,  Tennessee 37830, USA}

\author{Eric R. Hoglund}
\email{hoglunder@ornl.gov}
\affiliation{Center for Nanophase Materials Sciences, Oak Ridge National Laboratory, Oak Ridge,  Tennessee 37830, USA}

\date{\today}

\begin{abstract}
	Novel experiments in transmission electron microscopy often require the development of specialized imaging and spectroscopy states, such as for flexible momentum-resolved high energy-resolution spectroscopy. This task is complicated by the need to align and configure many different lenses, and by ambiguity around the locations of various imaging and diffraction planes. Here we develop a versatile digital twin for simulating the electron beam trajectory throughout an electron microscope, and we develop the calibration procedures required for accurate prediction of microscope states. This allows the user to quickly and easily determine the correct lens values for setting up new condenser or projector modes. Automatic procedures to change lens settings and measure the resulting changes can be used to close the loop. With the accelerating development of machine learning and artificial intelligence tools, we also believe a physically-informed model of the microscope can serve as a valuable tool for automated microscopy and AI/ML integration.
\end{abstract}

\maketitle

\newpage

\section{Introduction}
Due to the complexity of alignment and configuration of the many lenses inside the electron microscope, the development of new imaging and spectrometer modes remains out of reach for many users. A lack of clarity regarding the optical configuration of the microscope poses a challenge for new and experienced microscopists alike. The precise optical configuration for manufacturer-defined settings are also often poorly document (if at all). Meanwhile, key parameters such as the calibrations of camera lengths or collection angles are critical for accurate quantitative analysis of data, and the ability to freely configure these values and subsequently record the microscope’s true state is essential. Furthermore, the ability to set up an arbitrary microscope state can enable new and better experimentation. Finally, partially- or fully-automated microscopy \cite{Spurgeon2020,Kalinin2021,Roccapriore2022,Kalinin2026} can benefit from a physically-informed model of the microscope, allowing acquisition and exploration beyond predefined workflows.

To address these concerns, we have developed an instrument digital twin for simulating the electron beam trajectory throughout the column. Ray optics are used to simulate the electron trajectory from gun to detectors, with a realistic treatment of elements including round lenses, quadrapoles, dipoles, apertures, and prisms. Lens strengths can be read from various sources (e.g., from the microscope itself via APIs if the microscope manufacturer allows, or from microscope configuration files). We also develop a road map for the calibrations of lens strengths (e.g. finding focal lengths from nominal electrical current readings or values given in terms of percentages of a maximum strength). This is done either through sweeps of lens parameters (with beam sizes, rotation, and total intensity recorded from the Ronchigram camera or other detector) or based on values read across multiple microscope configurations. Similarly, only approximate knowledge of the physical position of elements is required, as fitting procedures allow for fine-tuning of guessed or physically-measured positions.

While usage of the term varies, it is often expected that a digital twin should have bidirectional communication with its physical analog \cite{Kritzinger2018,Jones2020,Fuller2020,Vanderhorn2021}. In our case, the digital twin can be updated from values read via the microscope’s API, which allows for constant visualization of lens and crossovers as the microscope configuration is changed. Conversely, assuming the manufacturer’s APIs also allow the setting of lens strengths, the digital twin can then be used to set the microscope configuration. Conventional minimization algorithms can also be applied to the digital twin (e.g. for determination of lens parameters for arbitrary camera length, or optimization of lens parameters for stability), enabling arbitrary control over the microscope configuration. 

Finally, the digital twin allows automatic recording of the microscope state, allowing reconstruction of the microscope during data analysis, as opposed to relying on nominal or previously-calibrated values recorded by the user.


\section{Methodology}
\subsection{Developing a toy microscope model}
We begin by developing a toy model for beam propagation within a microscope. Critically, calibration of the model is required to create a fully functional digital twin, however the uncalibrated toy model may still be useful for visualization and demonstration purposes \cite{Landers2023}. Calibration will be discussed in a later section. All code for this work is publicly available via our github repository: \href{https://github.com/sea-ecosystem/rayTEM/}{github.com/sea-ecosystem/rayTEM/}

The beam trajectory though the microscope is simulated with ray optics and the transfer matrix method. An individual ray is minimally tracked by its positions ($x$,$y$) and angles ($\theta$ corresponding to the angle in $x$ and $\phi$ corresponding to the angle in $y$). Our code also tracks additional terms in the matrix such as beam intensity and energy $E$ in order to allow dispersing of the beam based on energy (via a prism element). Additional attributes are also tracked independently: total beam current (in order to account for current-reducing apertures), position $z$ along the length of the beam, and beam rotation angle $R$ at each point. In this work we will focus on the effects of round lenses and apertures however, as these primarily determine the imaging states available. Propagation through an element (given by a matrix $M$) is found via matrix multiplication: $|x_2,\theta_2,y_2,\phi_2| = M \times |x_1,\theta_1,y_1,\phi_1|$. This method has been used extensively for light optics \cite{Brouwer1964,Hawkes1989}, and to a lesser degree electron optics \cite{Brown1983,Krivanek1991,Craven2017,Landers2023}, so we will only briefly discuss the specific matrix operations for the most common elements. 

Our matrix representation of each element is directly based on that given by Brown \cite{Brown1983}. We will only use the first-order approximation, but we note that it still captures advanced phenomena such as the rotation of the electron beam within a solenoid. In this model, definitions are also based on field strength $K$ as opposed to focal lengths $f$.


For a round lens or solenoid, focusing and rotation of the electron beam uses the following matrix:
\begin{equation}
	\mathcal{S} = \begin{vmatrix} 
		C^2 & \frac{1}{K}\ S \ C & S\ C & \frac{1}{K} \ S^2 \ \\ 
		-K\ S\ C & C^2 & -K\ S^2 & S\ C \ \\
		-S\ C & -\frac{1}{K}\ S^2 & C^2 & \frac{1}{K}\ S\ C \ \\
		K\ S^2 & -S\ C & -K\ S\ C & C^2 \\
	\end{vmatrix} 
	\label{eq:lens}
\end{equation}
where $K$ is the field strength, $C=cos(K\ L)$ for a lens of length $L$, and $S=sin(K\ L)$. We highlight the fact that for small $K\ L$, the small-angle approximation applies: $cos(K\ L) \approx 1$ and $sin(K\ L) \approx K\ L$. Focusing can then be approximated as $f^{-1} \approx K^2\ L$. This approximation has been explored previously \cite{Pierce1949}, and we include additional analysis and discussion in the Supplemental Material. This approximation also simplifies calibration of lenses (i.e., relating $K$ to physical parameters in the microscope such as lens length and electrical current), as will be discussed in a later section. This also reduces the above expression to the more common single-axis definition for a round lens with no beam rotation:
\begin{equation}
	\begin{vmatrix} 
		x_2 \\ \theta_{2}
	\end{vmatrix} = \begin{vmatrix} 
		1 & 0 \\ -\frac{1}{f} & 1 
	\end{vmatrix} \begin{vmatrix} 
		x_1 \\ \theta_1
	\end{vmatrix}
	\label{eq:lenssimple}
\end{equation}

$\mathcal{S}$ can also be separated into rotating and focusing terms: $\mathcal{S}_{total}=\mathcal{S}_{rot} \times \mathcal{S}_{focus}$, shown below:
\begin{equation}
	\mathcal{S}_{rot} \times \mathcal{S}_{focus} = \begin{vmatrix} 
		C & 0 & S & 0 \ \\ 
		0 & C & 0 & S \ \\
		-S & 0 & C & 0 \ \\
		0 & -S & 0 & C \ \\
	\end{vmatrix} 
	\begin{vmatrix} 
		C & \frac{1}{K}\ S & 0 & 0 \ \\ 
		-K\ S & C & 0 & 0 \ \\
		0 & 0 & C & \frac{1}{K}\ S \ \\
		0 & 0 & -K\ S & C \\
	\end{vmatrix}
	\label{eq:lensrotation}
\end{equation}
where a round lens rotates the beam simply by an angle of $\theta = -K\ L$. In our model, we use this latter definition, as it enables the use a rotating reference frame to simplify the analysis. If beam rotation is tracked separately, rays within the rotating reference frame are simply focused via $\mathcal{S}_{focus}$. 

A drift segment for free-space propagation is modeled via the following matrix:
\begin{equation}
	\mathscr{D}=
	\begin{vmatrix} 
		1 & L \\ 0 & 1 
	\end{vmatrix}
	\label{eq:drift}
\end{equation}
for propagation length $L$. 

There is no convenient matrix representation of an aperture, however the key behaviors of an aperture can be captured via a custom propagation function. An aperture limits the size of the beam, so in the first-order aberration-free case, we simply rescale the $x$, $y$, $\theta$ and $\phi$ values of all rays. An aperture also changes the total beam current by masking out a portion of the beam, so we track the total beam intensity before and after each element, and update the intensity based on the rescaling performed. A more rigorous calculation of beam current based on the masked area would be required if aberrations were included however, and while not used for this work, this functionality is available in our code. 

To construct a toy microscope model from individual components, we begin with the condenser round lenses, the Virtual Objective Aperture (VOA), a pair of objective round lenses, projector round lenses, and Ronchigram detector (a plane in $z$ along the beam path where image and diffraction planes, beam sizes, etc, will be recorded). This is simply defined as a series of element matrices (drift, lens, drift, and so on). The starting rays are defined by initial positions and angles $x,y,\theta,\phi$, and the resulting rays at various points can be calculated. 

While a physical microscope typically has more dipoles than round lenses, these are primarily used for steering the beam as required for alignment through lenses. We have therefore neglected dipoles from our initial model, as dipoles will not affect the positioning (along the length of the column) and magnification of the various image and diffraction planes. This is equivalent to using a laterally-varying reference frame, where the origin follows the center of the beam. Similarly, while quadrupoles and hexapoles are used for aberration correction or to reduce astigmatism \cite{Haider1998,Krivanek1991,Krivanek2008}, the first-order matrix representations of beam optics used here do not capture aberrations, so these are neglected. Optical assemblies (such as a probe corrector) which meaningfully affect the focusing of the beam are reduced to a single round lens with an effective focal length and effective position, or neglected in certain cases. 

Critical outputs from the model may include lens rotation, the locations of crossovers, and locations and magnification of various image planes and diffraction planes. Rotation is tracked by summing each lens' contribution, and crossovers can be found by calculating the extent of the farthest (radially) ray. Image planes and diffraction planes can be inferred based on starting ray positions and angles: two rays emitted from the same point will cross at an image plane, and two rays emitted at the same angle will cross at a diffraction plane.

\subsection{Calibrating the model as a digital microscope twin}
While the modeling discussed above may yield a toy model useful for visualization and educational purposes \cite{Landers2023}, the calibration of lenses is required to create a physically meaningful digital twin \cite{Craven2017,Kritzinger2018,Fuller2020,Jones2020,Vanderhorn2021}. In other words, nominal values read from the microscope (often in terms of electrical current ($I$) or shown as a percentage of the lens's maximum magnitude) must be mapped to field strengths $K$ used by the ray tracing model. Furthermore, the functional relationship between $I$ and $K$ may not be known. While it should theoretically be linear, it may not be across the full range of lens values: electronics may behave nonlinearly, or lens component materials may reach magnetic saturation. Similarly, lens positions and thicknesses ($L$) must be found. Various strategies for calibration of the microscope model are outlined below, with additional details available in the Supplemental Material. We will focus on the alignment of the projector lens system for each of the examples to follow. 

\subsubsection{Inference of focal lengths as a function of lens strength}
To calibrate the nominal lens strengths to focal lengths, it may seem intuitive to simply sweep each lens individually, recording parameters such as beam diameter on the Ronchigram camera. This may hint at the functional relationship between lens strength and focal length, however it will not adequately find the correct scaling factors (e.g. $K=A*I$) if more than one lens is active. An example of this is shown in the Supplemental Material, where it can be shown that for two lenses, the proportionality for beam diameter vs.\ first lens focal length depends on the focusing of the second lens. This can also be understood analytically via the equations representing the effective focal length $f_{1,2}$ for a two-lens system, which depend on each individual lens focal lengths $f_1$ and $f_2$ and the distance between lenses $d_{2}$ \cite{FultzHowe2013}:
\begin{equation}
	\frac{1}{f_{1,2}} = \frac{1}{f_1}+\frac{1}{f_2}+\frac{d_{2}}{f_1 f_2}
	\label{eq:2lens}
\end{equation}
The measurement of the varying beam diameter under perturbation of a single lens $f_1$ or $f_2$ therefore depends on the focal length of other lenses in the system. 

\begin{figure}
	\centering
	\includegraphics[width=.98\linewidth]{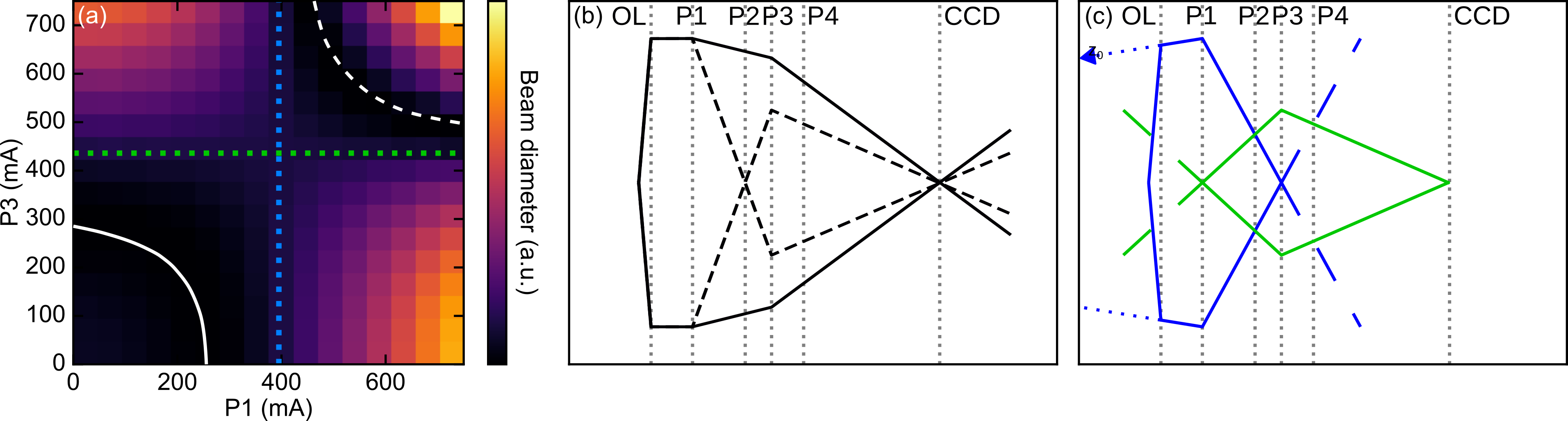}
	\caption{(a) Beam diameter is measured (shown as heatmap intensity) as a function of two projector lens strengths ($P1$ and $P3$). low- and high-current bands of minima (traced in white), correspond to focusing of the real-space probe to the CCD (ray diagram shown in (b)). Each lens has a critical current (vertical blue, horizontal green lines) where the beam diameter is insensitive to the other lens strength. These correspond to lens 1 ($I_{1,3\ critical}$) placing a crossover at lens 3 (a,d, blue), or lens 3 ($I_{3,1\ critical}$) projecting the CCD plane back to lens 1 (a,b, green). In both cases, Eq. \ref{eq:1lens} allow calculation of electrical lens current $I$ to the lens strength $K$}
	\label{fig:PL1vPL3}
\end{figure}

2D lens sweeps provide more information, however the scaling will again depend on the behavior of subsequent lenses. An example 2D sweep is shown in Fig. \ref{fig:PL1vPL3}.a, showing the beam diameter (heatmap intensity) as a function of the strength of projection lenses 1 and 3. Relevant ray paths are also shown in Fig. \ref{fig:PL1vPL3}.b,c for the scenarios discussed below. 

During normal operation, the objective lens forms a diffraction plane prior to the first projection lens. When sweeping the pair of projector lenses, two arcs of local minima can be observed, passing through two regimes. At low current for both lenses (Fig. \ref{fig:PL1vPL3}.a and b, solid white and black respectively), the focused STEM probe is refocused onto the detector. This defines an image plane. As both lens currents are increased, this plane is brought closer, a second diffraction plane appears, and a second image plane appears as well (Fig. \ref{fig:PL1vPL3}.a and b, dashed white and black respectively). Understanding these regimes is helpful for course calibration: if a given combination of lens values (e.g., a manufacturer-defined projection lens setting) yields an image plane or diffraction plane at the detector, it may not be clear in which regime this occurs. Sweeping up or down to explore where the next plane occurs may lend insight.

A surface fitting scheme may be able to replicate general features (arcs of minimum diameter, in approximately the correct locations) while obtaining incorrect calibration values however. 
One may also find discrepancies between fitted calibrations from sweeps of different lenses (1 vs 2, 1 vs 3, and so on). 

Instead, we note the vertical and horizontal lines of uniform intensity (uniform beam diameter) on the heatmaps (Fig. \ref{fig:PL1vPL3}.a, blue and green). In the scenario where lens 1 places a crossover at the plane of lens 3 (in this example), the beam diameter will be insensitive to the strength of lens 3 (Fig \ref{fig:PL1vPL3}.a,c, blue). Stated mathematically: $\frac{d\ diameter}{d\ I_{3}}=0$ if and only if lens 1 focuses to $z_{3}$, i.e, at a critical value of $I_{1}$. In the context of Eq. \ref{eq:2lens}, the effective focal length of the lens pair is insensitive to $f_2$ (i.e. $\frac{d\ f_{1,2}}{d\ f_2}=0$) only when lens 2 is placed at the image plane of lens 1 \cite{FultzHowe2013}. We thus define $I_{1,3\ critical}$ to be the value of $I_{1}$ at which $\frac{d\ diameter}{d\ I_{3}}=0$. Note that the value of $I_{1,3\ critical}$ will depend on the angle at which rays are entering lens 1 (or the location of the initial image plane). To construct an analytical representation of the lenses (allowing one to solve a system of equations for many configurations of multiple lenses), we define a virtual beam point source at $z_0$, where the axial rays leaving $z_0$ pass through the first lens ($z_1$ in this case) and refocus to the second ($z_3$).

Understanding that these critical values occur when the second lens is placed at the first lens's image plane, we can simplify the analytical expression to the single-lens formula: 
\begin{equation}
	\frac{1}{f} = \frac{1}{d_1}+\frac{1}{d_2}
	\label{eq:1lens}
\end{equation}
$f$ is the focal length, and $d_1$ and $d_2$ are the distances from the object (or initial crossover) to lens, and lens to image plane (subsequent crossover), respectively. This can also be derived from the simplified drift and lens expressions from Eq. \ref{eq:lenssimple} and \ref{eq:drift}. In the case of an already-convergent beam entering the lens, there is still a virtual source positioned after the lens, $d_1$ is negative, and the same expression holds. Focusing the beam from any lens $i$ at position $z_i$ to the plane of a given subsequent lens $j$ at position $z_j$ thus follows the expression: $f^{-1}=\Delta z_{virtual,i}^{-1}+\Delta z_{i,j}^{-1}$ (where $d_1$ and $d_2$ in Eq. \ref{eq:1lens} are replaced by the differences in positions, $\Delta z$).

We can also consider the inverse case, where the later lens $j$ is set such that the beam diameter is insensitive to a given preceding len $i$, i.e. $\frac{d\ diameter}{d\ I_{i}}=0$ at a critical lens $j$ current $I_{j,i\ critical}$ (Fig. \ref{fig:PL1vPL3}.a, horizontal green). The intuitive explanation for this relies on the back-propagation of rays from the Ronchigram detector, i.e. a ray traced from the center of the detector outwards is focused back by lens 3 to the plane of lens 1. The ray diagram for $I_{3,1\ critical}$ is shown in green in Fig \ref{fig:PL1vPL3}.c. These rays should thus follow the same relationship $f^{-1}=\Delta Z_{i,j}^{-1}+\Delta z_{j, detector}^{-1}$. Alternatively, one can think in terms of the projection of image planes: an image plane at lens $i$ is projected via lens $j$ onto the Ronchigram detector (and adjusting the strength of lens $i$ affects only the angles, not the positions, at the plane of lens $i$). A similar argument is made for apertures: when the size of an aperture is found to be insensitive to the following lens strength, this is indicative that the aperture plane is projected to the detector (or subsequent virtual crossover $z_{virtual}$), i.e., $f^{-1}=\Delta Z_{aperture,i}^{-1}+\Delta z_{i,detector}^{-1}$.





\begin{figure}
	\centering
	\includegraphics[width=.98\linewidth]{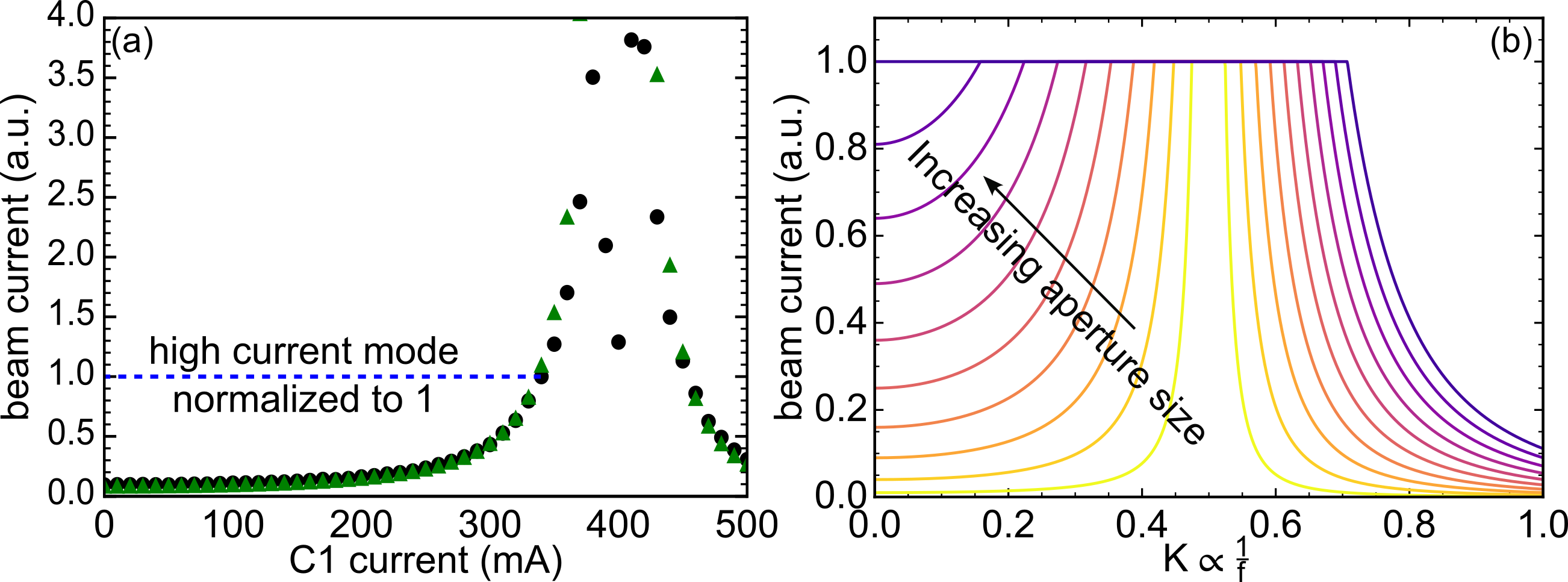}
	\caption{(a) crossovers through apertures can be found by measuring the total beam current while sweeping a preceeding lens, with the beam current vs C1 strength shown. (b) The trends seen in (a) are universal to any beam passing through an aperture, following the analytical expressions from Eqs. \ref{eq:aperture1} and \ref{eq:aperture2} for a range of aperture diameters. An asymptote exists at low current, approaching the zero lens strength current. at high-$K$, the focal position is brought in front of the aperture, and continuing to increase $K$ will continue to decrease the beam current.}
	\label{fig:C1Aperture}
\end{figure}

For a beam passing through an aperture, calibration of focal lengths to lens strengths can instead use the measured beam current as a function of lens strength. As the diameter of the beam entering the aperture expands, the beam current is reduced by the square of the ratio of beam diameter vs.\ aperture diameter (Eqs. \ref{eq:aperture1},\ref{eq:aperture2}). Fitting the beam current vs.\ lens current curve can therefore determine the focal strength at which the lens focused to the aperture plane, along with the aperture diameter relative to the initial beam diameter. The analytical expressions for these curves are shown below, derived from the simplified lens and drift matrices in Eq. \ref{eq:lenssimple} and Eq. \ref{eq:drift}, for the beam entering the lens at position $x_1$ and angle $\theta_1$. Raw and fitted data are shown in Fig \ref{fig:C1Aperture}.a, and example curves for a range of aperture sizes are shown in Fig \ref{fig:C1Aperture}.b. We use this analysis in our calibration of our condenser lenses (elements ordered C1/VOA/C2/C3).

\begin{equation}
	x_2 = x_1+L\ (\theta_1-\frac{x_1}{f}) 
	\label{eq:aperture1}
\end{equation}

\begin{equation}
	I_{beam} = \begin{cases}
		(x_2/r_{aperture})^2 & \mathrm{if}\ \ \ \ x_2 > r_{aperture} \\
		1 & \mathrm{else}
	\end{cases}
	\label{eq:aperture2}
\end{equation}

In all cases, a single measurement of a single $I_{critical}$ value provides one analytical lens equation (Eq. \ref{eq:1lens}). Provided enough measurements, the system of equations can be solved for all unknowns (lens calibrations between electrical current $I$ and field strength $K$, lens positions, and so on). Calculated positions of virtual image planes also means the beam angles entering or exiting the given section will be found, i.e. the ``effective'' position of the gun, required divergence/convergence from a gun lens, the position of pre-objective lens crossovers, and the beam divergence/convergence entering the projector section. 

This is only true for cases where the thin-lens approximation is valid however, i.e. when $K*L$ is sufficiently small. For the thick-lens case, there may still be a suitable thin-lens equivalent, however rigorous ray propagation (initializing the rays at a fixed plane and measuring where they reconverge) may be required depending on the accuracy required. During calibration of our microscope model, we neglect the thickness of the condenser lenses, as beam rotation is not a critical parameter of interest. For the projector section, we have found it helpful to begin calibration using the analytical math, then use the results as guesses for the full rigorous model-based fitting. 

In our Nion microscopes, we find a nearly-linear $K = A*I$ relationship in all cases (within the regimes at which lenses are typically driven). When transitioning between the analytical thin-lens approximation and fully-rigorous model-based fitting, the $f^{-1} \approx K^2\ L$ approximation (with initial assumed lens length $L$) means the lens length and calibration factor $A$ can be simultaneously updated, while largely maintaining the expected focal lengths $f$. 

Both upstream and downstream $I_{critical}$ strengths can also be found without the need for the full 2D lens sweeps. This is done by wobbling lenses, e.g. apply perturbations to lens 2 and find the strength for lens 1 at which the beam does not change diameter, finding $I_{1,2\ critical}$. 


Finally, we note that the calculation of the positions of diffraction planes is highly sensitive to the precise position and focal length of the objective lens. Calibration of $P1$-$P4$ using the above procedure will find the position of the image plane produced by the second half of the objective lens ($O2$) (assuming a fixed sample position), however the position of the first diffraction plane follows $\Delta z_{O2,diff}=\Delta z_{sample,O2}+f$. Unknown position and calibration of the objective lens may thus yield the correct image plane, but incorrect diffraction plane (which therefore affects the accuracy of all subsequent diffraction planes). 
To calibrate the strength and precise position of $O2$, the same procedure used before can be applied to the second objective lens: adjusting $O2$ while wobbling $P1$-$P4$ to find several $O2$ currents with known focal lengths. Alternatively, if two or more microscope states are known to have correctly-focused diffraction planes with a differing number of crossovers, then the position and focal length of $O2$ in the model can be adjusted until both known states yield a diffraction plane in identical position, while maintaining the position of the $OL$ image plane at $z_0$.

\subsubsection{Measuring lens rotation}
The aforementioned procedures should reasonably characterize the focal length, however the relation to lens strength also depends on lens length: $f^{-1} \approx K^2\ L$. Critically, the rotation of the beam depends linearly on both lens strength and length: $\theta =K\ L$, meaning measurement of lens rotation is critical for differentiating between the effects of field strength and lens length, and for the ability to predict lens rotation at arbitrary lens combinations. 

If a pure beam shift can be applied at each point in the 1D or 2D sweeps, then the movement of the beam on the detector can also be measured (e.g., changing center of mass). This can be done at the sample plane (via the scan coils, with descan turned off), or prior (via any pair of dipoles, such as those following the objective lens or even within the condenser section). In theory, there should be zero beam motion when a diffraction plane is placed on the Ronchigram detector, and the distance of motion should determine the magnification for an image plane. Similarly, the direction of motion should serve as a measurement of the rotation angle induced by the projector lenses. In practice, data obtained in this manner was usually too noisy to be of use. Small shifts may not yield large enough shifts on the Ronchigram to accurately quantify, and large shifts may introduce comatic aberration through the round lenses, complicating the measurement of the center-of-mass. 

Alternatively, the beam can be intentionally stigmated upstream, leading to ellipticity of the VOA, and the angle of the ellipse can be measured. This method yields acceptable data, however we have developed an additional technique below which appears to be slightly more robust. 

Our preferred method for measuring beam rotation involves acquiring a Ronchigram image of a defocused sample. For example, when our calibration sample (gold nanoparticles on a lacy carbon grid) is imaged at large defocus, the lacy pattern can be seen rotating as the round lenses are adjusted. Justifying the images to measure rotation is done by fitting the ellipse of the VOA, recentering (in case imperfect alignment results in movement of the VOA as each lens is adjusted), anisotropically scaling until the VOA is round (in case slight astigmatism yields ellipticity of the VOA), and then rotating until multiple Ronchigram images are aligned (quantified via the mean-squared deviation between images). As expected, the rotation is linear with strength, however an abrupt flip of 180 \degree is seen where the beam passes through a crossover (image is inverted, or mirrored in both $x$ and $y$). While this method is significantly more complex, it appears to be slightly more robost (a comparison between rotations found using the rotation of the ellipse vs.\ the rotation of the defocused Ronchigram image is available in the Supplemental Material). 

Using the focal lengths found above, and beam rotations found here, the calibration for both strength and length can be found. 

\begin{figure}
	\centering
	\includegraphics[width=1\linewidth]{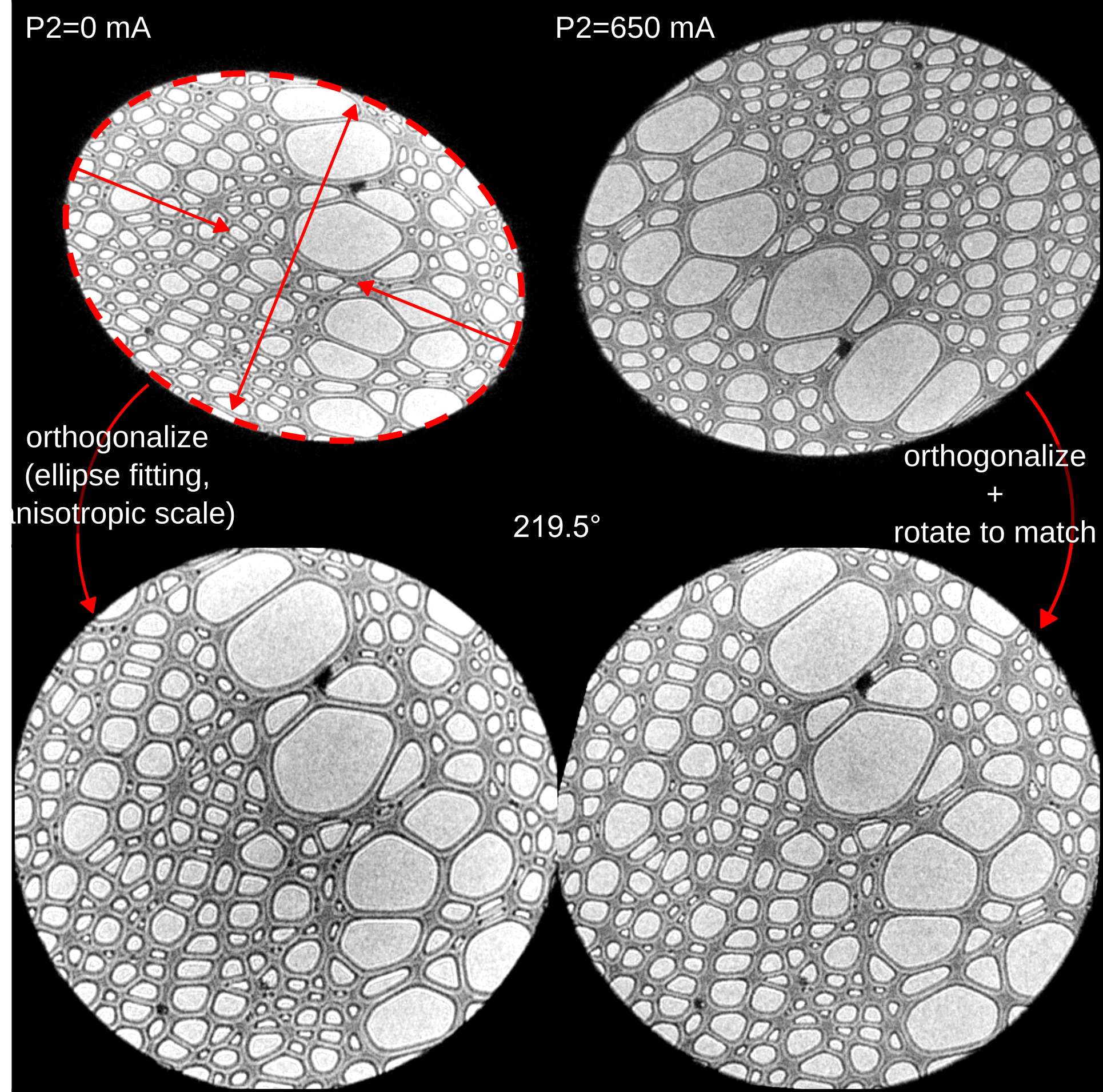}
	\caption{(a) Ronchigram images of lacy carbon at two lens strengths can be used to measure the relative rotation of the lens between these strengths}
	\label{fig:rotationalalignment}
\end{figure}

\section{Example use cases}
Once calibration parameters are found (translating nominal lens current $I$ to lens strength $K$ in the model, and fine-tuning lens positions and lengths), fitting can be performed to find the optimal lens configuration for a given set of requirements. 

In the condenser section, the beam current is predicted based on the beam diameter through the VOA, the focus at the sample plane is determined by the position of the pre- objective lens crossover, and the convergence angle is determined by the diameter of the beam entering the objective lens. To achieve a desired convergence angle, at a desired beam current, any minimization algorithm can be used to calculate $I_{C1}$, $I_{C2}$ and $I_{C3}$ to adjust beam current and convergence angle while maintaining focus. 

For the projector section, we truncate the model, initializing the rays at the sample plane. This avoids propagation of error through the condenser section, and the initialized rays can contain both axial and field rays for calculating the position of diffraction and image planes respectively. Magnification of a real-space image plane is calculated by the ratio of ray's starting and final position in $x$ or $y$, and the magnification of the diffraction plane is based on the starting ray angle departing the sample plane and final position (this is also referred to as ``camera length'', where the position of a diffracted beam is $x=\theta*l$ on the detector, for scattering angle $\theta$, after length $l$). The positions of image or diffraction planes along the length of the column is crucial for proper focusing of diffraction images, and coupling into a spectrometer \cite{Craven2017}. Finally, rotation of the beam between the objective lens and a detector may also be of interest, in order to rotationally align diffraction images. 

Before developing new projection modes, existing modes and their coupling to the spectrometer must first be understood. Interestingly, our reference setting does not yield a diffraction plane at the CCD, and the locations of post-projector diffraction planes are inconsistent between various pre-existing settings. While focus of the diffraction plane can be determined based the apparent sharpness of the VOA, it may be easier to symmetrically wobble the objective lenses (i.e. perturb the strength over and under focus) and observe the changes in the size of the Bragg disks. When the size of the Bragg disks changes symmetrically, the diffraction plane is focused to the detector.

First, we must recognize that the zero-loss beam in the spectrometer is an image plane along the energy-dispersive axis i.e., the zero-loss beam is focused to a point, capturing all electrons within the cone of the STEM probe, and diffracted STEM probes at all angles. All fixed lenses following the projector section therefore serve to project a post-projector image plane, and the location of this image plane is critical for coupling into the spectrometer. 

Slight defocus of the diffraction plane on the CCD may also only negligibly affect results. At points aside from a mathematically-ideal diffraction plane, diffracted disks may still appear. slight defocus (blurry edges of disks) may not be immediately apparent, and apertures applied at these planes might still effectively select the desired scattering angles. While the ideal scenario is projection of a diffraction plane onto the entrance aperture, this may not always be achievable. Loosening the criteria requiring a fixed post-projector diffraction plane position may allow greater apparent magnification or better control over diffraction pattern rotation in conjunction with magnification. 

Finally, we note that results may depend on the choice of minimization algorithm used. Basic gradient descent algorithms may have difficulty when moving between regimes, i.e. microscope states with additional or fewer crossovers than present in the guessed or initial configuration. This can be understood in the context of Fig. \ref{fig:PL1vPL3}.b. If one seeks to place a crossover at the CCD plane, the closest crossover may need to be moved farther before the next crossover appears. A pre-computed grid across many combinations of the lenses of interest (condenser lenses or projector lenses) may therefore be useful for finding the initial guesses to pass to a minimizer. 

\subsection{non-overlapping Bragg disks}
We begin with a basic example involving the condenser section. For a range of 4D-STEM experiments \cite{Ophus2019,Roccapriore2022}, it may be desirable to find the largest convergence angle which yields non-overlapping Bragg disks. This will maximize the spatial resolution attainable while still allowing measurement of things like strain and orientation mapping. 

Beginning from our 30 mrad reference setting, we use a least squares minimization scheme using the aforementioned criteria to calculate the condenser settings for a new convergence angle, and push the raw fitted lens strengths to the microscope. Due to slight error in the calibration of the lenses or slight non-linearity in the lenses, there may be deviation between nominal and actual resultant convergence angle, and the sample may be slightly out of focus. We measure the error in focus by adjusting the objective lens focus control to refocus the sample, avoiding possible hysteresis of the motion stage. We then return the objective lens focus to it's original value, and make minor adjustments to $I_{C3}$ to return the sample to focus, recording the correct (as opposed to nominal) lens value. Slight misalignment of the beam through the condenser lenses may also mean coupling into the probe corrector has changed, so we adjust this using the standard procedure (e.g. wobbling the probe corrector's last quadrupole and adjusting dipoles). Finally, we measure the convergence angle via the Ronchigram. We tabulate the error in focus, the nominal and actual values for $I_{C3}$ (indicative of the accuracy of our lens calibration), and nominal vs.\ actual convergence angles, in Table \ref{table:CL} below, for a few convergence angles. Lens calibrations appear to be accurate, with higher at higher lens strengths (far from the regime where the lenses were calibrated). Small error in calibration leads to large defocus (e.g. 0.4\% error yielding 245 nm in defocus) due to the high strength of the objective lens and small distance between objective lens and focal position within the pole piece. 

\begin{figure}
	\centering
	\includegraphics[width=.98\linewidth]{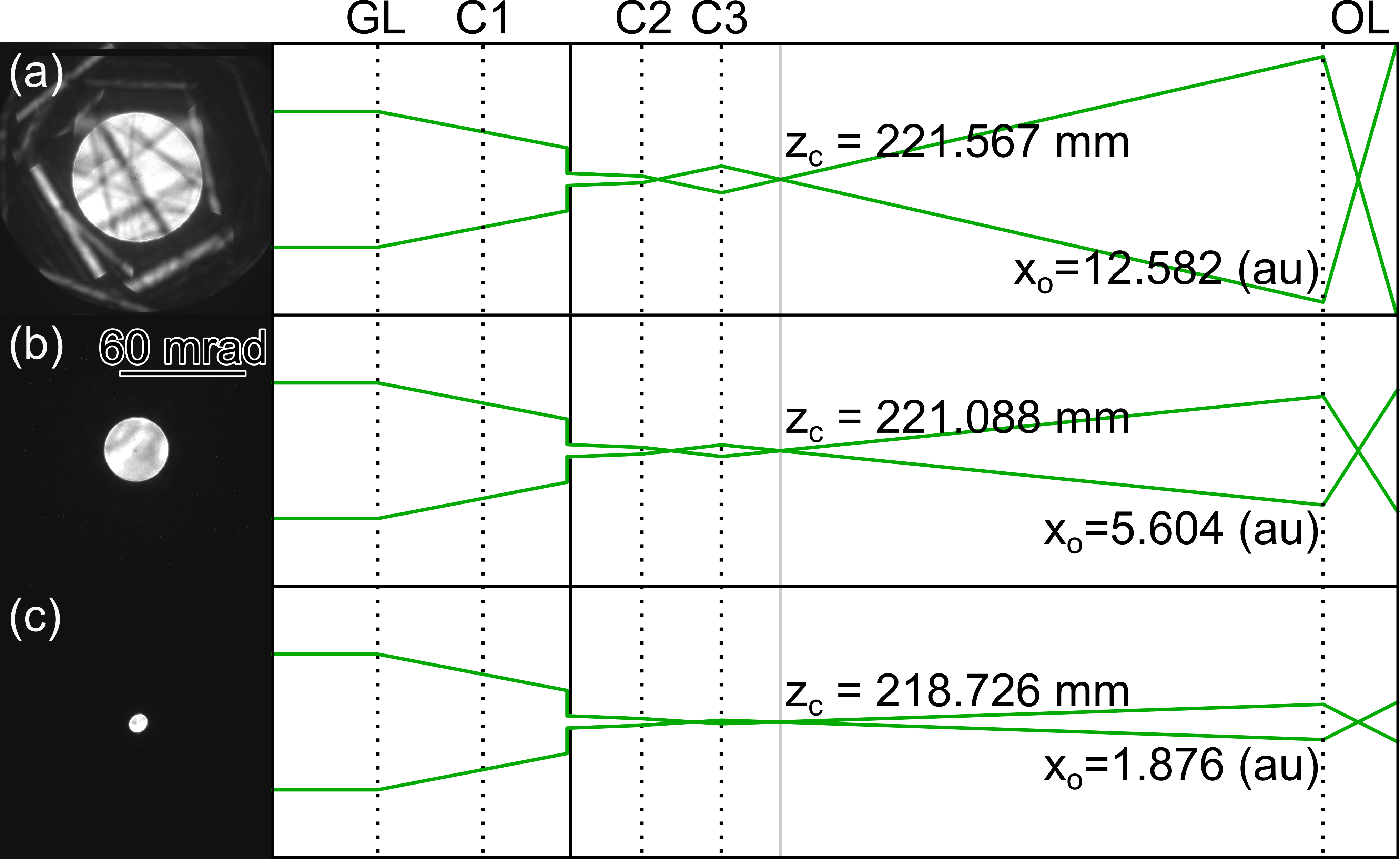}
	\caption{Ronchigram images and modeled microscope states for three convergence angles. These correspond to the lens parameters tabulated in Table \ref{table:CL}. Using the microscope model, arbitrary new states can be found: focus is controlled by the position of the post-$C3$ crossover ($z_c$), and we use the beam diameter entering the objective lens ($x_o$) as a proxy for the convergence angle (with proportionality dependent on OL strength).}
	\label{fig:CLdemo}
\end{figure}

	\begin{table*}
		\centering
		\begin{tabular}{|c|c|c|c|c|c|c|c|c|c|c|} 
			\hline
			& $C1$ & $C2$ & $C3$ (nominal) & $C3$ (actual) & $\Delta C3$ \% & $\Delta$ focus (nm) & mrad (modeled)  & mrad (actual) & $\Delta$ mrad & $\Delta$ mrad (\%) \\
			\hline
			a & 0 & 758.222 & 856.682        & 856.682       & -              & -                   & 30.947  & 30.947  & - & - \\
			b & 0 & 509.436	& 907.177        & 910.817       & 0.40 \%        & 245.55              & 13.802  & 14.592 	& -0.79 & -5.41\% \\
			c & 0 & 298.779 & 1092.379       & 1108.888      & 1.49 \%        & 812.67              & 4.6234  & 4.453   & 0.17 & 3.83\%  \\
			\hline
		\end{tabular}
		\caption{When the microscope is set to fitted lens values ($C3$ nominal), slight defocus may result, which can be corrected through minor adjustments to condenser lenses ($C3$ actual). The model predicts the convergence angle (mrad modeled) which is compared to that measured from the Ronchigram. row (a) is used as the reference, with the actual values for focus and convergence angle based on deviations from the reference. }
		\label{table:CL}
	\end{table*}

\subsection{Understanding nano-beam and TEM modes}

\begin{figure}
	\centering
	\includegraphics[width=.98\linewidth]{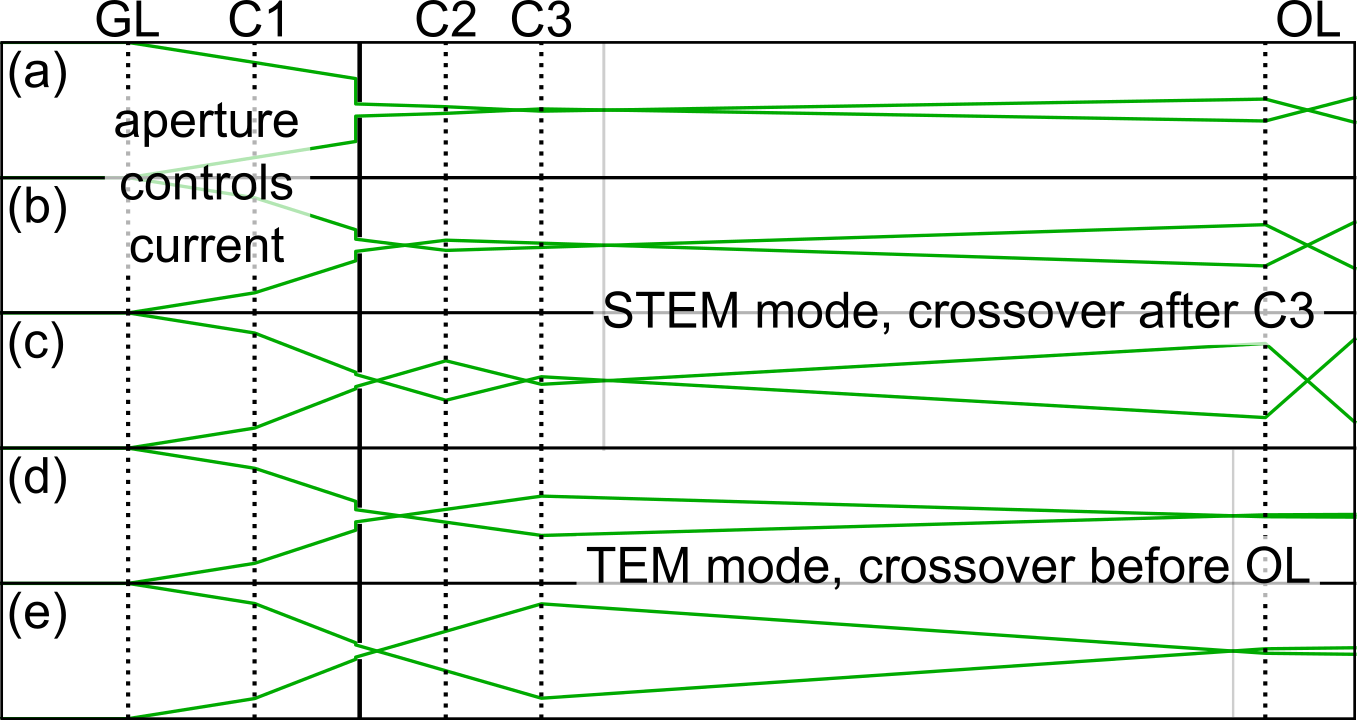}
	\caption{Various microscope states are modeled. (a,b,c) show low-, medium-, and high-current STEM modes: $C1$ is adjusted to control current through the aperture, while $C2$ and $C3$ are adjusted to maintain the position of the post-$C3$ crossover. The additional crossover between $C2$ and $C3$ (shown in a,c, absent in b) may or may not be required depending on the incident angle resulting from $C1$. TEM modes (d,e) are obtained by positioning the crossover immediately before the objective lens.}
	\label{fig:lowAlpha}
\end{figure}

The microscope model also aides in our understanding the lower limits of convergence angle and the various regimes available. To set the focus of a STEM probe, we maintain the position of the crossover after lens $C3$, meaning we may try to minimize the size of the beam entering $C3$ via $C1$ and/or $C2$, and use $C3$ to maintain the position of the crossover. In contrast, a TEM mode or K\"{o}hler illumination mode \cite{Benner1994} would position the crossover much closer to the objective lens. Both cases are shown modeled in Figure \ref{fig:lowAlpha}. In the case of the STEM modes in which $C2$ places a crossover immediately before $C3$ (Fig. \ref{fig:lowAlpha}.a,c), the lower limit for convergence angle depends on the maximum strength for $C3$ (minimum focal length, or maximum incoming convergence angle allowed) and the precision of how closely one can place the crossover to $C3$. Using lenses farther away ($C1$ and $C2$ instead of $C2$ and $C3$) to achieve the post-$C3$ crossover therefore yields a smaller convergence angle before reaching the limits of lens strengths. 

\subsection{Arbitrary control over projection modes}
With a calibrated model of the projection system lenses, we can calculate and set arbitrary (within the limits of the lenses) camera lengths and Ronchigram rotation. As in our example exploring convergence angles previously, we use the model to calculate projector lens strengths according to our desired camera length and rotation, and push the raw fitted values to the microscope. If the correct post-objective diffraction plane is known, any setting which maintains the position of this plane should maintain focus, and we found very little deviation in diffraction plane focus (within the range which could be ascertained through small wobbles of the objective lens). Positions of Bragg disks are measured however, allowing measurement of the rotation and magnification of the diffraction image. In general, since our rotations were measured accurately and the rotational calibration was applied following lens calibration, the rotational accuracy of the model is quite good (within a few degrees). Compounding error through multiple lenses or post-specimen aberrations \cite{hoglund2024,lupini2025} may lead to moderate error in the predicted camera length however (up to 10\% in extreme cases). These results are tabulated in Table \ref{table:PL} below. Rotation and deviation in magnification are calculated relative to our re-focused reference setting. In some cases, a 180$\degree$ inversion is expected, in cases where an additional beam crossover occurs.  The upper limit on camera length depends on the ability to maintain focus of the diffraction plane to the detector or aperture plane, and on the maximum lens currents available. 



	\begin{table*}
		\centering
		\begin{tabular}{|c|c|c|c|c|c|c|c|c|c|c|} 
			\hline
			& $P1$ & $P2$ & $P3$ & $P4$ & 
			$\theta$ (modeled) (\degree) & $\theta$ (actual) (\degree) & $\Delta \theta$ (\degree) & 
			$l$ (modeled) (mm) & $l$ (actual) (mm) & $\Delta l$ (\%) \\  \hline
			a & 451.0   & 0.      & 0.      & 560.5   &       -   &     -     &    -  &  20   &   -   & -   \\ 
			b & 370.750 & 387.238 & 0.0     & 547.517 &     -17.3 &   -18.8   &  1.5  & 9.20  & 9.19  & 0.13  \\	
			c & 541.665 & 1075.68 & 547.74  & 563.9   &     75.9  &   71.7    &  4.2  & 29.8  & 31.8  & -6.3 \\ 
			d & 551.264 & 999.999 & 180.706 & 610.204 &     99.1  &    96.5   &  2.6  & 48.7  & 53.6  & -9.2 \\ 
			\hline
		\end{tabular}
		\caption{When the microscope is set to fitted lens values ($P1-P4$), the expected and measured rotation angle and camera length may differ. Measurements are taken based on Ronchigram images collected for each microscope state. Each of these states is shown in Fig. \ref{fig:PLstates}. row (a) is used as the reference, with the actual values for rotation and camera length based on deviations from the reference. Note that a 180$\degree$ inversion occurs for (c) and (d).}
		\label{table:PL}
	\end{table*}

\begin{figure}
	\centering
	\includegraphics[width=.98\linewidth]{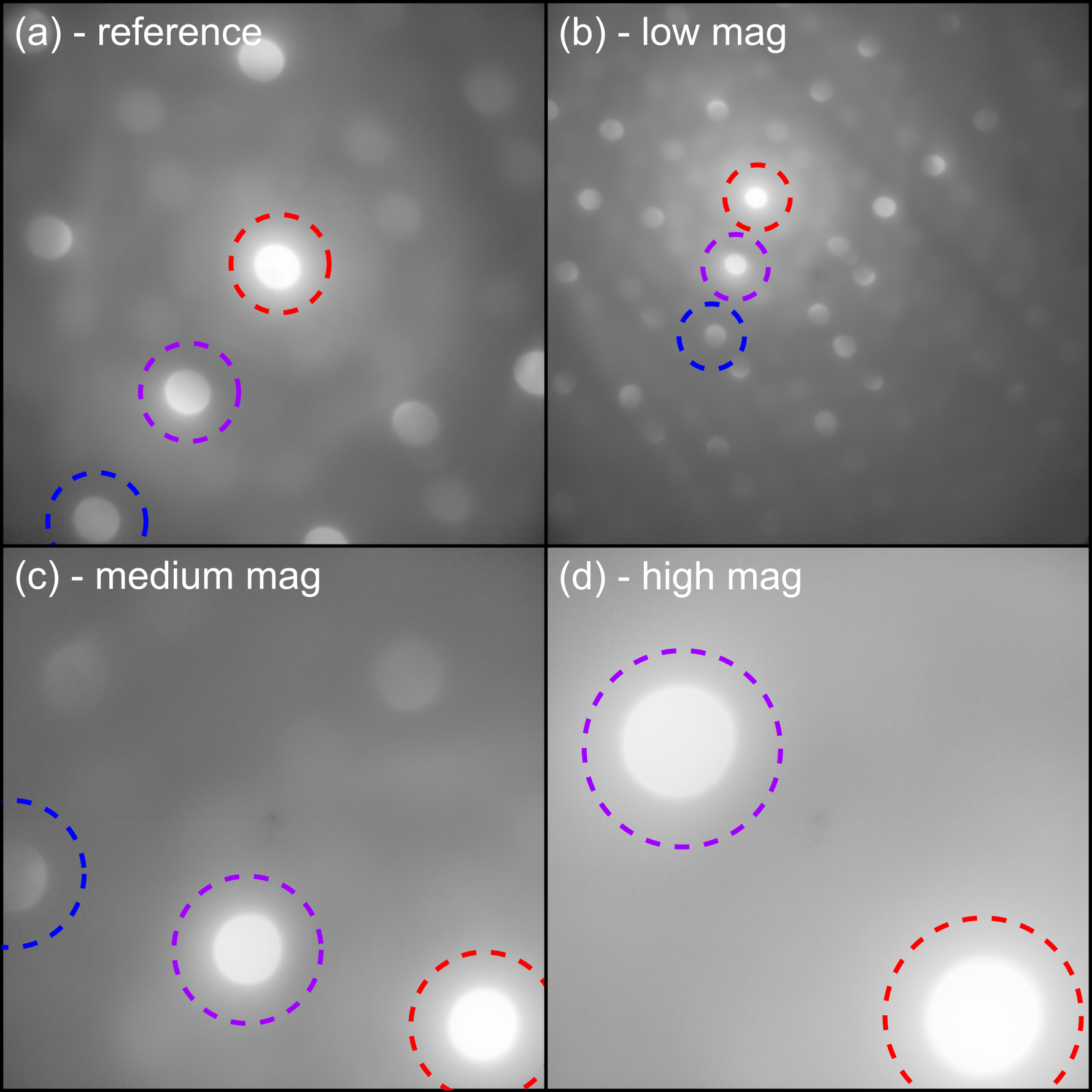}
	\caption{The projector section states tabulated in table \ref{table:PL} are shown. Common Bragg disks are circled, to enable visualization of camera length and rotation.}
	\label{fig:PLstates}
\end{figure}







\subsection{Momentum-resolved EELS}
In a momentum-resolved EELS experiment, a slot aperture is used, installed with the length perpendicular to the dispersion axis of the spectrometer \cite{Pettit1975,Plotkinswing2020,Qi2021,Elgvin2025}. A diffraction plane is placed at the aperture plane (meaning data is collected in momentum along the slit direction), and the beam is focused and dispersed in energy in the perpendicular direction. The extent of data collected in momentum thus depends on the magnification of the diffraction image at the aperture plane (or camera length). The magnification required to collect out to, for example, 1 Brillouin zone may therefore differ depending on the sample. The momentum direction also depends on the rotation of the diffraction image relative to the slit. The collection angle in the dispersive direction (width of the slit, relative to the Bragg disk diameter) also partially determines the energy resolution \cite{Krivanek2013,Craven2017}, so it is considered ideal to match the Bragg disk diameter to the slit width. Finally, depending on the level of monochromation, beam current may be low and acquisition times long, meaning the highest beam current possible through the VOA is desired. 

\begin{figure}
\centering
\includegraphics[width=.98\linewidth]{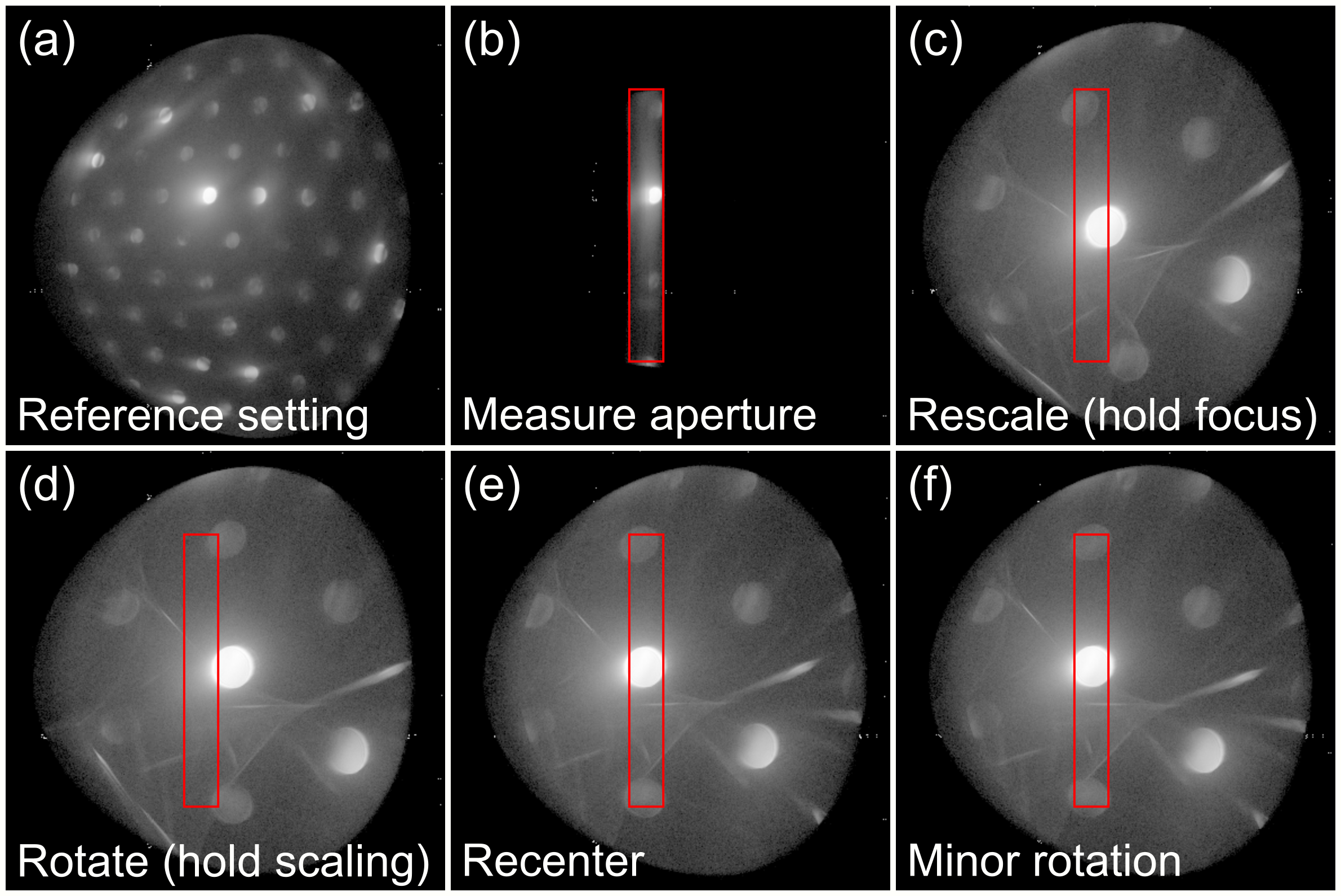}
\caption{Momentum resolved EELS requires simultaneously adjusting the magnification of the diffraction image (camera length), rotation to align the reciprocal directions of interest with the aperture, while maintaining position of the image plane within the microscope. These steps are performed the ray-tracing model and fitting to identify and set the appropriate lens parameters.}
\label{fig:PLdemo}
\end{figure}

Our criteria for the condenser section are therefore: low convergence angle set to approximately match the Bragg disk diameter to slot aperture width, beam current maximized, with focus maintained at the sample plane. Our criteria for the projector section are: diffraction plane positioned at the aperture plane, diffraction plane magnification (or camera length) set to scale the distance to the first Bragg disks to the aperture length, and diffraction plane rotated such that the first Bragg disks are along the slit direction. A step-by-step iterative procedure is demonstrated in Figure \ref{fig:PLdemo}, however with sufficiently accurate calibration and sufficiently robust fitting algorithms, this could be performed automatically \cite{Koster1992}. 

\section{Conclusion}
In this work, we have developed a simple physics-informed digital twin to model the behavior of a STEM microscope. The model is broken down into simplified and verifiable sections (e.g., condenser, objective, and projector sections), which are universally applicable to a variety of microscopes from any manufacturer. We also offer robust procedures for calibration, allowing inference of the positions of lenses, and translation between lens current to focal length. Optical elements are simulated as simple matrix operations, meaning calculations are fast and easily extensible to higher-order to account for aberrations. While a toy microscope model is useful for visualization and educational purposes, calibration of said model offers predictive power, offering the ability to calculate arbitrary (within the limits of maximum lens strengths) microscope configuration, which can enable better measurements. Finally, we believe our realistic physics-informed model can help with AI integration to electron microscopy. AI-driven workflows could yield semi- or fully-automated data acquisition across a range of microscope configurations \cite{Spurgeon2020,Kalinin2021,Roccapriore2022,Pratiush2025,Kalinin2026}, and AI-integration could simplify troubleshooting and assist in training. 

\section{Acknowledgments}

T.W.P \& E.R.H. acknowledge the support of the U.S. Department of Energy, Office of Basic Energy Sciences (DOE-BES), Division of Materials Sciences and Engineering under contract  ERKCK47.

Electron microscopy was conducted at the Center for Nanophase Materials Sciences (CNMS), which is a US Department of Energy, Office of Science User Facility at Oak Ridge National Laboratory.

Notice: This manuscript has been authored by UT-Battelle, LLC, under contract DE-AC05-00OR22725 with the US Department of Energy (DOE). The US government retains and the publisher, by accepting the article for publication, acknowledges that the US government retains a nonexclusive, paid-up, irrevocable, worldwide license to publish or reproduce the published form of this manuscript, or allow others to do so, for US government purposes. DOE will provide public access to these results of federally sponsored research in accordance with the DOE Public Access Plan (https://www.energy.gov/doe-public-access-plan).

\bibliographystyle{unsrt}
\bibliography{Refs.bib}

\end{document}


\renewcommand{\thefigure}{S\arabic{figure}}
\renewcommand{\thetable}{S\arabic{table}}
\renewcommand{\theequation}{S\arabic{equation}}

\linespread{1.6}\selectfont{} 
\preprint{JRN/123-ABC}

\title{Development of a versatile generalized digital twin for TEM and STEM - Supplemental Material}

\author{Thomas W. Pfeifer}
\email{pfeifertw@ornl.gov}
\affiliation{Center for Nanophase Materials Sciences, Oak Ridge National Laboratory, Oak Ridge,  Tennessee 37830, USA}

\author{Andrew R. Lupini}
\email{arl1000@ornl.gov}
\affiliation{Center for Nanophase Materials Sciences, Oak Ridge National Laboratory, Oak Ridge,  Tennessee 37830, USA}

\author{Eric R. Hoglund}
\email{hoglunder@ornl.gov}
\affiliation{Center for Nanophase Materials Sciences, Oak Ridge National Laboratory, Oak Ridge,  Tennessee 37830, USA}

\date{\today}

\maketitle

\newpage

\section{\label{sec:level1} Supplemental Material}

\subsection{Code availability}
Our ray-tracing code package is available at \hyperlink{https://github.com/sea-ecosystem/rayTEM}{https://github.com/sea-ecosystem/rayTEM}. A microscope is modeled via a ``Microscope'' object, containing a series of ``MicroscopeSection'' objects, each of which contain a series of ``Element'' objects. Starting rays are either provided, or generated from a ``Source'' object. Rays are propagated through the appropriate element types, such as ``Drift'', ``Lens'', ``Aperture'', and so on. Convenience functionality is also available, e.g. movement of lenses by automatically adjusting the length of preceding and subsequent drift elements, calculation of image and diffraction planes, and pre-defined error functions for use in a minimization/fitting scheme. 

No calibrated microscope models are provided online or with this manuscript, however we are happy to share calibrated models \textit{with vendor permission}. Calibrated models are also unlikely to be directly transferrable to other systems, due to subtle variation in lens configuration, arrangement, and variations in lenses (even across microscopes made by the same manufacturer). Instead, we hope the calibration procedures outlined here are sufficient for the owner of a microscope to assemble their own model and perform their own calibration. One should know approximate lens positions, and should have the ability to reasonably control lenses, in order to find focal length ($I_{critical}$) conditions. Generalized code for fitting is available on the github page, and the authors can provide guidance for any special cases or where customized fitting required. 


\subsection{Basic example of the inadequacy of 1D lens sweeps}
In the main manuscript, we suggested that 1D sweeps of individual lenses, with measurements of parameters such as beam diameter on the ronchigram, are inadequate. This is due to the unknown scaling of any subsequent or prior lenses, which should be apparent from the two-lens focusing equation:

\begin{equation}
    \frac{1}{f_{1,2}} = \frac{1}{f_1}+\frac{1}{f_2}+\frac{d_{2}}{f_1 f_2}
   \label{eq:2lens}
\end{equation}
 
or can be shown visually:

\begin{figure}
    \centering
    \includegraphics[width=0.95\linewidth]{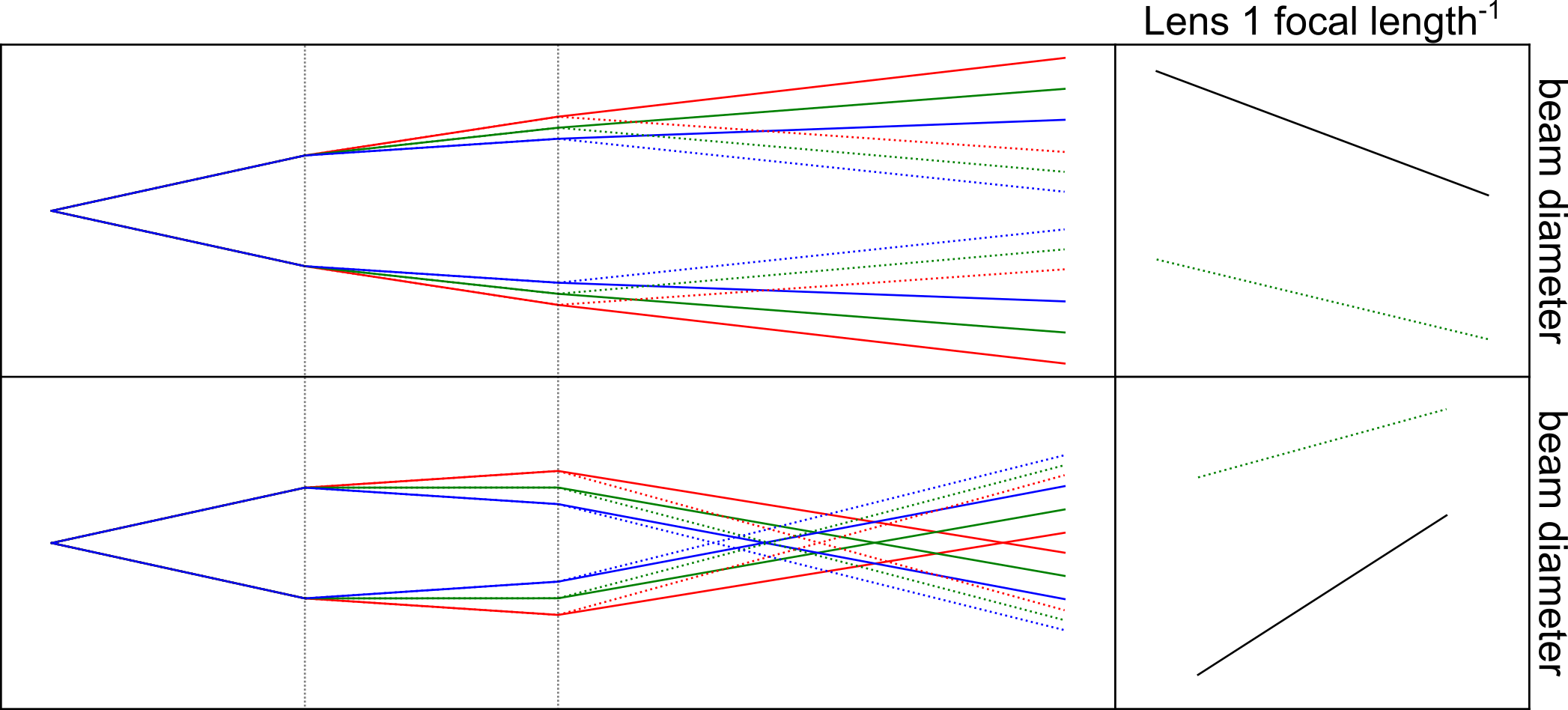}
    \caption{In a two-lens system, linear changes to strength yield a linear change in beam diameter, however preceeding or subsequent lenses affect the slope of these linear changes, meaning sweeps of individual lenses are insufficient for calibration.}
    \label{fig:demo1D}
\end{figure}

\subsection{Assessing the accuracy of the $f^{-1}\approx K^2 L$ assumption}
In the main manuscript, we use the small angle approximation: if $K\ L$ is small, then $sin(K\ L) \approx K\ L$, $cos(K\ L) \approx 1$, and $f^{-1}\approx K^2 L$. This approximation offers a significant simplification to the analysis, and allows the easy use of expressions such as $f^{-1}=d_1^{-1}+d_2^{-1}$ in systems of equations to solve for lens current calibration factors to go between lens current $I$ and field strength $K$ as appearing in the matrix expressions for round lenses. Here, we evaluate the accuracy of this assumption. 

Considering in 2D for simplicity (or one axis in the rotating reference frame), we have the simplified lens matrix:
\begin{equation}
    \begin{vmatrix} 
    x_2 \\ \theta_{2}
    \end{vmatrix} = 
    \begin{vmatrix} 
   C & \frac{1}{K}\ S  \ \\ 
   -K\ S & C  \ \\
   \end{vmatrix} 
    \begin{vmatrix} 
    x_1 \\ \theta_1
   \end{vmatrix}
   \label{eq:lensrotation}
\end{equation}

where $C=cos(K\ L_l)$ and $S=sin(K\ L_l)$, for lens strength $K$ and lens length $L_l$. For parallel starting rays at position $x_1$ and $\theta_1=0$, we seek to focus this ray to $x_3=0$ via a subsequent drift segment:

\begin{equation}
    \begin{vmatrix} 
    x_3 \\ \theta_{3}
    \end{vmatrix} = 
    \begin{vmatrix} 
   1 & L_d  \ \\ 
   0 & 1  \ \\
   \end{vmatrix} 
    \begin{vmatrix} 
    x_2 \\ \theta_2
   \end{vmatrix}
   \label{eq:lensrotation}
\end{equation}

for a drift of length $L_d$. The combined matrix is: 



\begin{equation}
    \begin{vmatrix} 
    x_3 \\ \theta_{3}
    \end{vmatrix} = 
    \begin{vmatrix} 
   1 & L_d  \ \\ 
   0 & 1  \ \\
   \end{vmatrix} 
 \begin{vmatrix} 
   C & \frac{1}{K}\ S  \ \\ 
   -K\ S & C  \ \\
   \end{vmatrix} 
    \begin{vmatrix} 
    x_1 \\ \theta_1
   \end{vmatrix} = 
   \begin{vmatrix} 
   C-L_d\ K\ S & \frac{1}{K}\ S+L_d\ C  \ \\ 
   -K\ S & C  \ \\
   \end{vmatrix} 
       \begin{vmatrix} 
    x_1 \\ \theta_1
   \end{vmatrix}
   \label{eq:lensrotation}
\end{equation}

and we seek $x_3=0$, or:

\begin{equation}
C-L_d\ K\ S= cos(K\ L_l)-L_d\ K\ sin(K\ L_l)=0
\end{equation}

Recall for a thin enough lens where $C=1$ and $S=K\ L$, this reduces to $1-L_d\ K^2 L_l=0$, which satisfied our approximation: $f^{-1}=L_d^{-1}=K^2\ L$. For moderate $L_l$ however, $L_d = C\ S^{-1}\ K^{-1}$, and the focal length (including the forwards propagation of the beam through the thickness of the lens, length $L$) if $f_{actual}=C\ S^{-1}\ K^{-1}+L$. In order to represent arbitrary $K$, $f$, $L$, we have plotted the percent deviation between $f_{approximate}=K^{-2} L^{-1}$ and $f_{actual}=C\ S^{-1}\ K^{-1}+L$ ($percent\ deviation = (f_{actual}-f_{approximate})/f_{actual}$) as a function of the ratio of lens length to focal distance ($L/f_{actual}$). The error in the approximation exceeds 5\% when $L/f$ reaches 7.5\%. When high lens rotation is observed, thick lenses are expected, and the approximation should not be used for large lens strengths where focal lengths will be smaller. 

\begin{figure}
    \centering
    \includegraphics[width=0.95\linewidth]{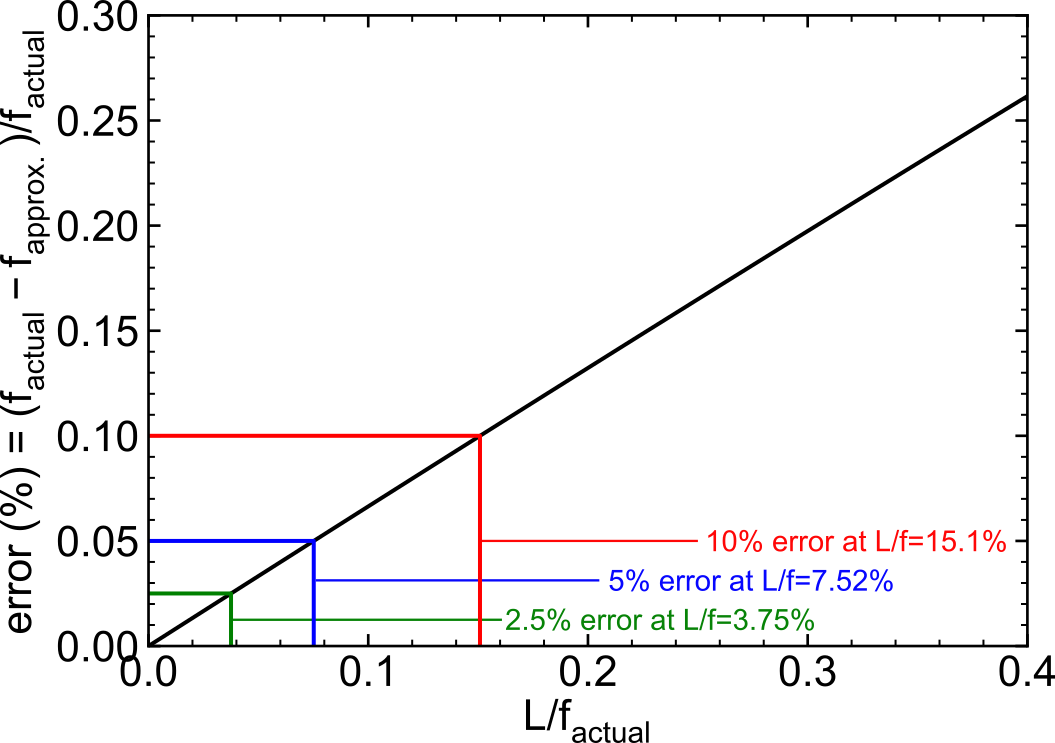}
    \caption{Deviation between the true focal length and the approximated focal length ($f^{-1}\approx K^2 L$, based on the small angle approximation) is plotted}
    \label{fig:focallength}
\end{figure}

\subsection{Measuring rotation via a defocused ronchigram image vs ellipse rotation}
In the main manuscript, we discussed two methods for measuring the beam rotation imparted by the projector lenses. While we do not claim one method is objectively better than the other, we do find slightly increased noise in our measured rotation when using the ellipse angle. The introduction of arbitrary 90 $\degree$ changes in angle (due to the ambiguity in ellipse orientation) also affects the analysis, whereas the image matching algorithm is significantly more complicated. 

\begin{figure}
    \centering
    \includegraphics[width=0.95\linewidth]{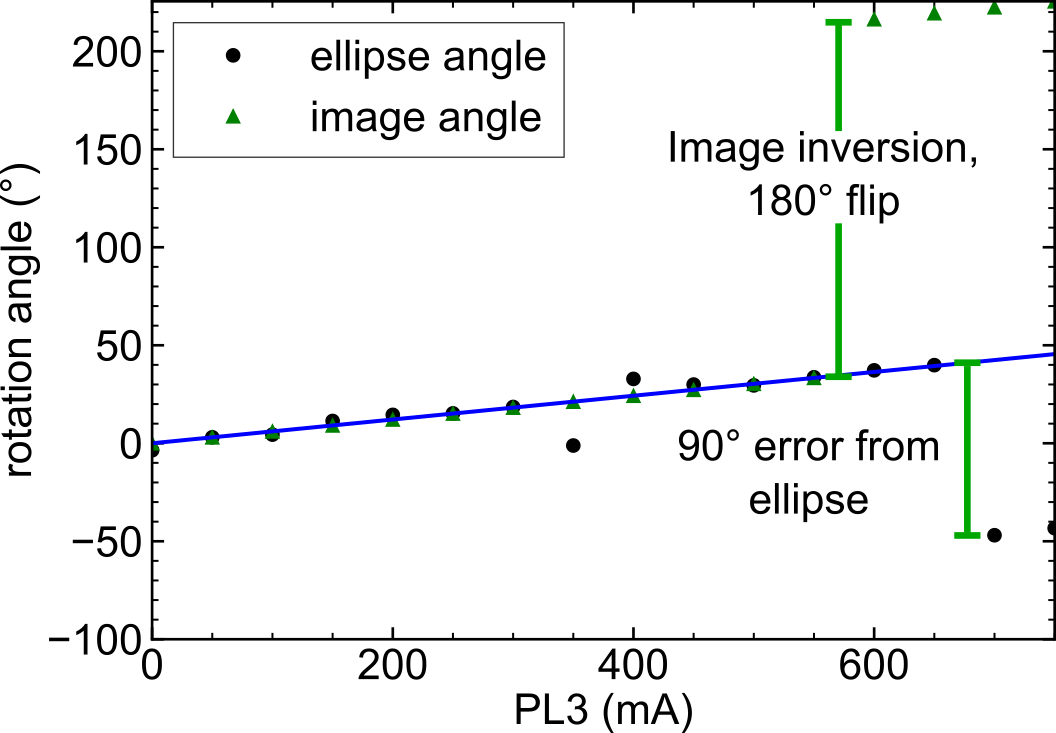}
    \caption{Beam rotation can be measured via the rotation of the ellipse produced by a stigmatic beam (black), however this can result in 90 $\degree$ error due to ambiguity in ellipse orientation. We instead use the rotation of a defocused ronchigram image (green), where the apparent rotation of the defocused sample can be tracked.}
    \label{fig:rotplot}
\end{figure}

\bibliographystyle{unsrt}
\bibliography{Refs.bib}